\documentstyle[prl,aps,epsfig]{revtex}
\newcommand{\KET}[1]{{| #1 \rangle}}

\newcommand{\cuoq}[0]{\mathrm{CuO}_{4} }

\newcommand{\commento}[1]{} 
 
\newcommand{\be}{\begin{equation}}
\newcommand{\ee}{\end{equation}}
\newcommand{\beq}{\begin{eqnarray}}
\newcommand{\eeq}{\end{eqnarray}}

\newcommand{\cu}[0]{\mathrm{CuO}_{4} }

\begin{document}
    
\def\gC{\mbox{\boldmath $C$}}
\def\gZ{\mbox{\boldmath $Z$}}
\def\gR{\mbox{\boldmath $R$}}
\def\gN{\mbox{\boldmath $N$}}
\def\ua{\uparrow}
\def\da{\downarrow}
\def\a{\alpha}
\def\b{\beta}
\def\g{\gamma}
\def\G{\Gamma}
\def\d{\delta}
\def\D{\Delta}
\def\e{\epsilon}
\def\ve{\varepsilon}
\def\z{\zeta}
\def\h{\eta}
\def\th{\theta}
\def\k{\kappa}
\def\l{\lambda}
\def\L{\Lambda}
\def\m{\mu}
\def\n{\nu}
\def\x{\xi}
\def\X{\Xi}
\def\p{\pi}
\def\P{\Pi}
\def\r{\rho}
\def\s{\sigma}
\def\S{\Sigma}
\def\t{\tau}
\def\f{\phi}
\def\vf{\varphi}
\def\F{\Phi}
\def\c{\chi}
\def\w{\omega}
\def\W{\Omega}
\def\Q{\Psi}
\def\q{\psi}
\def\de{\partial}
\def\inf{\infty}
\def\ra{\rightarrow}
\def\bra{\langle}
\def\ket{\rangle}

\draft

\twocolumn[\hsize\textwidth\columnwidth\hsize\csname
@twocolumnfalse\endcsname

\widetext

\title{Repulsion-Sustained Supercurrent and Flux Quantization  in  Rings 
of Symmetric Hubbard Clusters}

\author{Agnese Callegari, Michele Cini,  Enrico Perfetto and Gianluca 
Stefanucci}
\address{Istituto Nazionale per la Fisica della Materia, Dipartimento di Fisica,\\
Universita' di Roma Tor Vergata, Via della Ricerca Scientifica, 1-00133\\
Roma, Italy}

\maketitle

\begin{abstract}

We test the  response  to a threading magnetic field of rings of 5-site  
$C_{4v}$-symmetric repulsive Hubbard clusters
connected by weak intercell links; each 5-site  unit has the topology of a 
CuO$_{4}$ cluster and a repulsive interaction is included on every site.
In a numerical study of  the three-unit ring with 8 particles, we take 
advantage of a novel
 exact-diagonalization technique which can be generally applied 
to many-fermion problems. For O-O hopping we find  Superconducting Flux 
Quantization (SFQ), but for purely Cu-Cu links 
 bound pair propagation is hindered by symmetry.
The results agree with $W=0$ pairing theory.

\end{abstract}

]

\narrowtext

{\small 

The repulsive Hubbard ring in the presence of a magnetic flux has 
been studied by several authors\cite{kusmartsev1}\cite{schofield}\cite{yufowler}\cite{nakano}.
An anomalous Aharonov-Bohm (AB) effect\cite{ba} with ground-state energy 
oscillations versus flux $\phi$ having a period shorter than the foundamental one 
 $\f_{0}=hc/e$ was reported, but no sign of superconductivity 
was found. 
The rings that we consider in this paper have nodes which consist  of 5-site units with a CuO$_{4}$ topology.
In the following we shall refer to the central site of each node  as Cu and to the four 
external sites as O just to distinguish their position in the unit 
cell, even if we acknowlegde that the present model may be too idealized 
to have much  direct relevance to high-$T_{c}$ cuprates. The 5-site cluster 
will be also called $\cu$.  Below we 
show numerical solutions of such a model that clearly show 
superconducting pair hopping if the total number of particles is 
$2|\L|+2p$ where $|\L|$ is the number of units and $0<p<|\L|$; in particular, 
once a magnetic field is switched on into the ring, SFQ is unambiguously observed. 
We note that in our model no superconducting response is obtained 
with less than 2 particles per 5-site unit;  the 
Zhang-Rice picture\cite{zr} for the two-dimensional $d-p$  
model does not represent superconducting pairs and for the present small 
system  we need to explore 
a scenario with a slightly larger density of particles.  By replacing 
CuO$_{4}$ by larger units one can  model other ranges of filling 
fraction.

The repulsive Hubbard Hamiltonian has {\em two-body} singlet eigenstates without double 
occupation\cite{cibal1}\cite{cibal2}\cite{cibal3}\cite{cibal4} called 
$W=0$ pairs.  Such solutions are also allowed in the fully symmetric clusters ${\cal C}$.
In the {\em many-body} ground state these pairs get dressed and 
bound, and this is signaled by  $\D_{{\cal C}}(N)<0$ where $\D_{{\cal C}}(N)=E^{(0)}_{{\cal C}}(N)+
E^{(0)}_{{\cal C}}(N-2)-2E^{(0)}_{{\cal C}}(N-1)$; $E^{(0)}_{{\cal C}}(N)$ 
is  the interacting ground state energy of  the cluster ${\cal C}$ 
with $N$ particles.  By means of a non-perturbative canonical 
transformation\cite{EPJB1999}\cite{SSC1999}, it can also be shown that 
$\D_{{\cal C}}(N)<0$ is due to an attractive effective interaction and 
at weak coupling  $|\D_{{\cal C}}(N)|$ is just the 
binding energy of the pair. The extension of the theory to the full 
plane was also put forth in Ref.\cite{EPJB1999}. 
The $C_{4v}$ symmetric 5-site cluster is the smallest one where the 
$W=0$ pairing mechanism works. We have already described 
$W=0$ pairing  in great detail as a function of the one-body and interaction 
parameters on all sites; the study was extended to larger clusters 
too\cite{cibal3}\cite{EPJB2000}.

We are using  CuO$_{4}$ as the unit just for the sake of 
simplicity, but  the $W=0$ mechanism produces bound pairs at 
different fillings for larger clusters\cite{cibal4} too. In order to 
simplify the analysis,  here 
we neglect the O-O hopping term within each unit and the only nonvanishing hopping matrix 
elements are those between an O site and the central Cu site; they 
are all equal to $t$.
In the total Hamiltonian  
\begin{equation} 
H_{\rm tot} = H_{0}+H_{\t},
\label{hlattice}
\end{equation}
$ H_{0}$ describes the units    and $H_{\t}$ the hopping between 
first-neighbor units.
\begin{eqnarray}
H_{0}=\sum_{\a=1}^{|\L|} [ t \sum_{i\s}( d^{\dag}_{\a \s}p_{\a, i\s}+
p_{\a, i\s}^{\dag}d_{\a\s})+\quad\quad\quad\nonumber \\ 
U(\hat{n}^{(d)}_{\a \ua} \hat{n}^{(d)}_{\a \da}+\sum_{i}\hat{n}^{
(p)}_{\a, i \ua}\hat{n}^{(p)}_{\a, i\da}
)  ]     
\label{senzatau} 
\end{eqnarray}
where $p^{\dag}_{\a, i\s}$ is the  creation operator 
onto the O $i=1,..,4$ of the $\a$-th unit and so on.
The point symmetry group of $H_{0}$ includes $S_{4}^{|\L|}$, 
with $|\L|$  the number of nodes; in this report we consider 
$|\L|=2$ and  $|\L|=3$. 

We take  $H_{\tau}$   invariant 
under the $S_{4}$  subgroup of $S_{4}^{|\L|}$, although a square 
symmetry would be enough to preserve the $\D_{\mathrm{CuO}_{4}}(4)<0$ property.
We analyzed alternative models for this term.

First,we studied   hopping between corresponding O sites, i.e. 
connecting the $i$-th O site of the 
$\a$-th unit to move towards the  $i$-th O site of the 
$\b$-th unit with hopping integral 
$\t_{\a\b}\equiv|\t_{\a\b}|e^{i\th_{\a\b}}$: 
\begin{equation}
H_{\tau}=\sum_{\a=\b\pm 1}\sum_{i\s}
\left[\t_{\a\b} p_{\a,i\s}^{\dag}p_{\b,i\s}+{\mathrm h.c.}\right]\; .
\label{htau}
\end{equation} 
In numerical work we assumed small $|\tau_{\a\b}| \ll 
|\Delta_{\cu}(4)|$. This is necessary because the clusters that we can diagonalize 
explicitly are small and a large $|\t|$ with a flux of the order of a 
fluxon would produce a strong  perturbation on the ground state 
multiplet. The same flux in a large system would indeed be 
perturbative even for $|\t| \sim t$.

For $N=2|\L|$ and $\tau_{\a\b}\equiv 0$, the unique  ground state consists 
of 2 particles in each CuO$_{4}$ unit. We assume a  total number 
of
$N=2|\L|+2p$ particles. When $U/t$ is such that 
$\D_{\mathrm{CuO}_{4}}(4)<0$, each of the $p$  added pairs prefers to lie on a single 
$\cu$  and for $N=2| \L | + 2p$ the unperturbed ground state is 
$2^{p}$$\times$${|\L|}\choose{p}$ times degenerate 
(since $^{1}{\cal E}$ has dimension 2).

We   exactly diagonalize the $|\L|=2$ and $|\L|=3$ ring Hamiltonian. 
The numerical work is  made easy by the powerful 
{\em Spin-Disentangled}  technique, 
which we briefly introduced recently\cite{JOPC2002}, but deserves a 
fuller illustration. To the best of our 
knowledge, it was not invented earlier, which is somewhat 
surprising, being a very general method. 

We let  $M_{\ua}+M_{\da}=N$ where $M_{\s}$ is the number of  particles 
of spin $\s$; $\{|\f_{\a\s}\ket\}$ is a real orthonormal basis, that 
is, each vector is a homogeneous polynomial in the $p^{\dag}$ and $d^{\dag}$ 
of degree $M_{\s}$ acting on the vacuum. We write the 
ground state wave function in the form 
\begin{equation}
|\Psi\rangle=\sum_{\alpha \beta}
L_{\alpha \beta}|\phi_{\alpha \uparrow}\rangle 
\otimes |\phi_{\beta \downarrow} \rangle 
\label{lali}
\end{equation}
which shows how the $\ua$ and $\da$ configurations are entangled. The 
particles of one spin are treated as the ``bath'' for those of the 
opposite spin: this form also enters the proof of a famous theorem by Lieb\cite{lieb}.
In Eq.(\ref{lali})   $L_{\a\b}$ is a 
$m_{\ua}\times m_{\da}$ rectangular matrix with 
$m_{\s}$=${5|\L|}\choose{M_{\s}}$. 
We let  $K_{\s}$ denote the kinetic energy $m_{\s}\times m_{\s}$ 
square matrix of $H_{\rm tot}$ in the basis $\{|\f_{\a\s}\ket\}$, 
and $N^{(\s)}_s$ the spin-$\s$ occupation number matrix at site 
$s$ in the same basis ($N^{(\s)}_s$ is a  symmetric matrix since the 
$|\f_{\a\s}\ket$'s are real). Then, $L$ is acted upon by the Hamiltonian 
$H_{\rm tot}$ according to the rule 
\begin{equation}
H_{\rm tot}[L]=[K_{\ua}L + L K_{\da}]+
U \sum_{s}  N^{(\ua)}_s L N^{(\da)}_s\;.
\label{lieb3}
\end{equation}
In particular for $M_{\ua}=M_{\da}$ ($S_{z}=0$ sector) 
it holds $K_{\ua}=K_{\da}$ and $N^{(\ua)}_s=N^{(\da)}_s$. Thus, the action 
of $H$ is obtained in a spin-disentangled way. 
The generality of the method 
is not spoiled by the fact that it is fastest in the $S_{z}=0$ 
sector, because it is useful provided that the spins are not totally 
lined up; on the other hand, $S_{z}=0$ can always be assumed,
as long as the Hamiltonian is $SU(2)$ invariant.

For illustration, consider the Hubbard model with two sites $a$ and 
$b$ and two electrons (H$_{2}$ molecule) each in the $\f_{a}$ or 
$\f_{b}$ orbital. The intersite hopping is $t$ 
and the on-site repulsion $U$. In the standard method, one 
sets up basis vectors for the $S_{z}=0$ sector
\begin{eqnarray}
|\q_{1}\ket=|\f_{a\ua}\ket\otimes |\f_{a\da}\ket,
\quad\quad
|\q_{2}\ket=|\f_{a\ua}\ket\otimes |\f_{b\da}\ket,
\nonumber\\ 
|\q_{3}\ket=|\f_{b\ua}\ket\otimes |\f_{a\da}\ket,
\quad\quad
|\q_{4}\ket=|\f_{b\ua}\ket\otimes |\f_{b\da}\ket,
\nonumber
\end{eqnarray}
One then looks for eigenstates (three singlets and one triplet)
\begin{equation}
|\Psi\ket=\sum_{i=1}^{4} \psi_{i}|\q_{i}\ket
\label{dard}
\end{equation}
of the Hamiltonian 
\begin{equation}
H_{{\rm H}_{2}}=\left(\begin{array}{rrrr}
U & t & t & 0 \\
t & 0 & 0 & t \\
t & 0 & 0 & t \\
0 & t & t & U \end{array}\right).\label{stand}
\end{equation}
Insted of  working with $4\times 4$ matrices, we can cope with $2\times 
2$ by the spin-disentangled method using the form in Eq.(\ref{lali}) with 
\begin{eqnarray*}
L=\left(\begin{array}{rr}
\psi_{1} & \psi_{2}\\
\psi_{3} & \psi_{4}  \end{array}\right),
\quad\quad K_{\s}=\left(\begin{array}{rr}
0 & t\\
t & 0 \end{array}\right), \nonumber \\
N_{a}^{(\s)}=\left(\begin{array}{rr}
1 & 0\\
0 & 0  \end{array}\right),
\quad\quad 
N_{b}^{(\s)}=\left(\begin{array}{rr}
0 & 0\\
0 & 1  \end{array}\right).\\
\end{eqnarray*}
Using Eq.(\ref{lieb3}), one finds 
\begin{equation}
H_{{\rm H}_{2}}|\Psi\rangle=\sum_{\alpha=a,b}\sum_{\beta=a,b}
(H_{{\rm H}_{2}}[L])_{\alpha \beta}|\phi_{\alpha \uparrow}\rangle 
\otimes |\phi_{\beta \downarrow} \rangle 
\label{hlali}
\end{equation}
with
\begin{equation}
H_{{\rm H}_{2}}[L]=\left(\begin{array}{cc}
U\psi_{1}+t(\psi_{2}+\psi_{3}) & t(\psi_{1}+\psi_{4})\\
t(\psi_{1}+\psi_{4}) & U\psi_{4}+t(\psi_{2}+\psi_{3}) 
\end{array}\right).\\
\end{equation}
The reader can readily verify that this is the same as 
applying $H_{{\rm H}_{2}}$ in the form of Eq.(\ref{stand}) to the standard wave 
function in Eq.(\ref{dard})
and then casting the result in the form of Eq.(\ref{lali}).
Since we can apply $H_{{\rm H}_{2}}$ we can also diagonalize it.
The advantage of working with $2 \times 2$ rather than $4 \times 4$ 
matrices for the H$_{2}$ toy model is ridiculous, but it grows with the size of the 
problem and in the  $|\L|=3$ ring case it is spectacular.
In the $S_{z}=0$ sector for $|\L|$=3 the size of the problem is 
1863225 and the storage of the Hamiltonian matrix requires  much space; 
by this device, we can work with matrices whose dimensions is the 
square root of those of the Hilbert space: $1365 \times 1365$  matrices 
solve the $1863225 \times 1863225$ problem, and are not even required to be 
sparse.

Here we have implemented this method for the Hubbard Hamiltonian. 
We emphasize, however, that this approach will be generally useful for 
the many-fermion problem, even with a realistic Coulomb interaction, 
which can be suitably discretized.  
 Starting from a trial wave function of the form in Eq.(\ref{lali}) we can 
avoid computing the Hamiltonian matrix, since its operation is given 
by Eq.(\ref{lieb3}). Each new application of the Hamiltonian takes 
us to a new Lanczos {\em site} and we can proceed by generating a Lanczos {\em chain}. 
To this end we need  to orthogonalize to the previous {\em sites} by the 
scalar product  given by $\bra \Q_{1}|\Q_{2} \ket = 
{\mathrm Tr}(L_{1}^{\dag}L_{2})$.
In this way  we put  the Hamiltonian matrix in a tri-diagonal form. This method 
is well suited since  we are mainly interested in  the 
low-lying part of the spectrum. 
A severe numerical instability sets in when the chain exceedes a few 
tens of {\em sites}, {\em i.e.}  well before the Lanczos method 
converges. Therefore we use 
repeated two-site chains alternated with moderate-size ones. 

In the basis of the sites (of the original cluster) the 
occupation matrices $N^{(\s)}_{s}$ are diagonal with elements equal to 
0 or 1, simplifying the calculation of the interaction term.
Moreover, in choosing the  trial wave function for small $\tau$
we  take full advantage from 
our knowledge of the $S_{4}$ irrep of the $\tau=0$ ground state. This 
speeds the calculation  by a factor of the order of 2 or 3 compared to a random 
starting state (or even more, if $U$ is large). 
Typically, starting from a  $\tau=0$ ground state for the three-unit  
ring, 24 short  Lanczos {\em chains} 
were enough to obtain a roughly correct energy and a 20-{\em 
site chain}  achieved an accurate eigenvalue and an already stabilized  
eigenvector. In limiting cases when the results could 
be checked against analytic ones, using double precision routines an 
accuracy better than 12 significant digits was readily obtained. 

The two-unit ring ($|\L|=2$) does not allow to insert a flux; the reason is that 
each unit is at the left {\em and} at the right of the other, and one 
cannot tell which is the clockwise motion. Results on the spectrum of 
such a system will be reported in a more comprehensive paper.  Here, 
for the three-unit ring ($|\L|=3$) we focus 
on the case ${U \over t} <34$ (which grants $\Delta_{\cuoq}(4)<0$) and $p=1$ (total number of 
particles $2|\L|+2p=8$). 

The  inter-unit hopping 
$\tau$ between the O sites breaks the symmetry  group 
$C_{3v}\otimes S_{4}^{3}$ into $C_{3v} \otimes 
S_{4}$ for real $\tau$; in a magnetic field (complex $\tau$), this 
further breaks into  $C_3 \otimes S_{4}$. More explicitly, let us label the ground state multiplet components
by the crystal 
momentum $ 2\p\hbar k/3$ where  $k$
 is an integer. Real $\tau$  separates  $k=0$ 
from  the 
$k=1$ and $k=2$ subspace  of $C_{3}$ , but cannot split 
$k=1$ and 2 because they belong to the degenerate irrep of $C_{3v}$;
this degeneracy
 is resolved by complex 
$\tau$  , when we insert a 
magnetic flux $\phi$ by  $\tau=|\tau| e^{i\theta}$,
$\theta=\frac{2 \pi}{3}(\phi/\phi_0)$.

The   persistent diamagnetic 
currents carried by bound pairs screen the magnetic flux.
We calculated  the expectation value  of the total current 
operator\cite{kohn} 
\begin{equation}
\hat{I} = c \frac{\de H_{\rm tot}}{\de \phi}= \frac{e}{\hbar |\Lambda|}
\sum_{i,\a,\s}i (\tau\, p_{\a+1,i\sigma}^{\dagger} p_{\a,i\sigma} - 
\tau^{*} p_{\a,i\sigma}^{\dagger} p_{\a+1,i\sigma} )
\label{current}
\end{equation}
as a function of the flux and versus $k$; this current operator yields a gauge
invariant average $I$. We note incidentally that by expanding $\hat{I}$ in Eq.(\ref{current}) in 
powers of $\f$ near $\f=0$ one may identify  paramagnetic and  
diamagnetic contributions with the zeroth and the first order terms 
respectively\cite{dagotto}, although such a splitting is not gauge 
invariant. 

The numerical results for $I$ are reported in 
Fig.(\ref{plot4}); they also convey direct evidence of the SFQ 
since  according to 
the Hellmann-Feynman theorem the current is proportional to the flux derivative 
of the ground-state energy. At $\phi=0$ the ground state has $k=0.$ 
Near $\f=0$ the system generates a 
diamagnetic current which screens the threaded magnetic field; 
$\f=0$  is a local minimum of the ground state energy, which
grows quadratically in 
$\phi$ (diamagnetic behaviour).
However, when $\phi$ exceedes a critical value $\sim \phi_{0}/4$, a 
symmetry change of the ground state occurs due to a level crossing 
 to $k=2$.
This corresponds to a sharp 
discontinuity of the current which suddenly changes sign; thus, for 
stronger flux   the current 
enhances the external field until, at 
$\f=\f_{0}/2$ the current vanishes again, signaling the energy 
minimum at half fluxon. This anti-screening of the field beyond the 
level crossing is interesting but the paradox is easily understood.
Indeed, like at $\f=0$, the eigenfuctions at $\f=\f_{0}/2$
may be choosen real, since $H_{\t}(\f=\f_{0}/2)$ is obtained from 
$H_{\t}(\f=0)$ by reversing the sign of four O-O $\t$  bonds connecting 
two nearest neighbours units. Thus, near $\f_{0}/2$ the magnetic flux 
can  still be considered as  a small perturbation, but 
the  unperturbed real inter-unit 
 Hamiltonian is $H_{\t}(\f=\f_{0}/2)$;  the current normally
 screens this perturbation. The vanishing of the current at 
$\phi=\phi_{0}/2$, when the ground 
state energy is in a new minimum belonging to the $k=2$ subspace, also 
marks the  restoring  of the time reversal invariance,  like 
in the BCS theory\cite{lipa}. Indeed, at 
$\phi=\phi_{0}/2$ the symmetry group is $\tilde{C}_{3v}\otimes 
S_{4}$  where  $\tilde{C}_{3v}$ is isomorphous to
$C_{3v}$ (reflections $\s$ are replaced by $\s g$, where $g$ is a suitable 
gauge transformation). This feature was also  found in other 
geometries\cite{EPJB1999}\cite{EPJB2000}.
Finally,  a level crossing at a critical value $\sim 3 \phi_{0}/4$ takes 
to the $k=1$ ground state and to the minimum at $\phi = \phi_{0}$; at 
this point the flux can be gauged away taking us back to $\phi =0, k=0$.
We numerically verified that $E_{k=2}(\phi)=E_{k=2}(\phi_{0}-\phi)$, 
$E_{k=0}(\phi)=E_{k=1}(\phi_{0}-\phi)$ and 
$E_{k=1}(\phi)=E_{k=0}(\phi_{0}-\phi)$, where $E_{k}(\f)$ is the 
ground state energy in the $k$ sector.
Thus, the dressed $W=0$ pair 
screens the vector potential like one particle with an effective charge 
$e^{\ast}=2e$ does. At both minima of $E^{(0)}(\f)$ we have 
computed $\D_{3-{\rm unit}}(8)\approx -10^{-2} t$.  
Here, the half-integer AB effect is actually SFQ.
 From Fig.(\ref{plot4}) we see that the maximum value of the 
diamagnetic current is of the order of $1\div 10$ nano Ampere if $t=1$ 
eV and the ratio ${I \over  \f/\f_{0}} \approx e|\t|/h$ near $\f=0$. 
The present results are in line with the expectations of the analytic 
theory that we have presented elsewhere\cite{JOPC2002}. As foreseen, 
this pattern disappears when $U/t \rightarrow 0.$

\begin{figure}[H]
\begin{center}
        \epsfig{figure=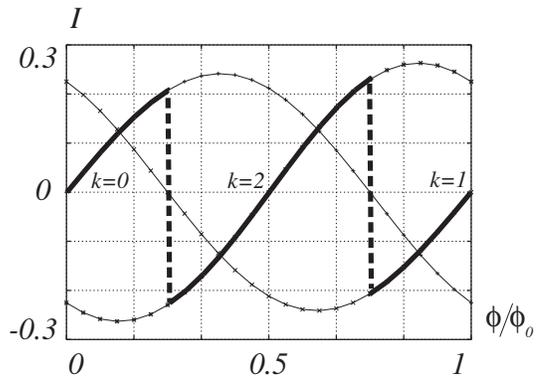,width=7cm}
        \caption{\footnotesize{Total current for the three-$\cuoq$ ring, as 
a function of the magnetic flux. Here $U=5\ 
t$, $|\tau|=0.001 \, t$. The current is in units of $e |\tau|/h$.
The thick line marks the ground state current.}}
\label{plot4}
\end{center}
\end{figure}

Next,  we report the effects of an inter-unit 
hopping $\tau_{\rm Cu}$ between Cu sites only; this does not 
break the $S_{4}^{|\Lambda |}$ symmetry and therefore its consequences 
on pair propagation and SFQ are drastically different.
In order to 
study the propagation of a bound pair we again assume the total number of 
particles $2|\L|+2=8$. The full system threaded by the flux has a 
$C_{3}\otimes S_4^3$ symmetry because the O sites are 
not involved in the inter-unit Hamiltonian. 
We computed the ground state energy versus flux for 
the case $\Delta_{\cuoq}(4)<0$; 
  $\tau_{\rm Cu}$ produces much smaller effects than $\t$ for 
$|\tau_{\rm Cu}|\ll |\Delta_{\cuoq}(4)|$.  
Therefore we are free to take the inter-unit hopping as large as 
$|\tau_{\rm Cu}|= \; 0.1 t$ without causing a drastic change;  
the dependence of the ground state energy is still weak, see Fig.(\ref{plot7}).
At $\phi=0$, the correction\cite{numeri} due to $|\tau_{\rm Cu}|$ to ground 
state energy is only $\sim 10^{-3} t$. However, the most spectacular 
change is that  the system now behaves like a paramagnet. 

\begin{figure}[H]
\begin{center}
        \epsfig{figure=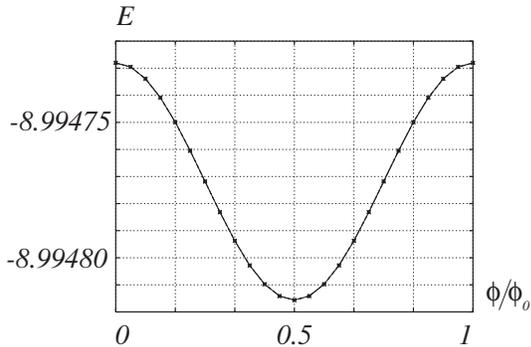,width=7cm}
        \caption{\footnotesize{ Ground state energy $E$ of 
	the three-unit ring in units of $t$, as 
a function of the concatenated magnetic flux. $E$ is $k$-independent 
(see text). Here $U=5\ 
t$, $|\tau_{\rm Cu}|=0.1 t$.}}
\label{plot7}
\end{center}
\end{figure}

The reason of this unusual behavior is the {\em local} symmetry, 
which forbids 
any  flux-induced 
splitting of the three $k$ levels; the effective mass is infinite and
 the $W=0$ pair, however bound, is 
localized.  Indeed the $S_{4}$ label of each 
$\cu$ unit is 
a good quantum number. No SFQ is observed because the screening 
of the magnetic field by the bound pair 
is forbidden. The small correction to the ground state energy 
comes from a second-order process. Suppose we prepare the system  {\em e.g.}
in  an unperturbed  state
$\KET{4,2,2}$, that is , with 4 particles in the first $\cu$ and 2 in 
the others;  the bound pair is localized on the first
cluster. The evolution of the wave packet  involves virtual 
states $\KET{3,3,2}$ and $\KET{3,2,3}$  which are obtained when a  
particle  
jumps to the nearby clusters. This particle must be 
in totalsymmetric state, because the irreps of each unit must be 
conserved. However, such a   process occurs with a tiny amplitude 
because the energy misfit is severe. This is particularly clear at weak coupling,
when we can speak in terms of orbitals, and 
the lowest-energy $A_{1}$ particle of the first cluster must hop to 
antibonding  $A_{1}$ orbitals of the nearby clusters; the amplitude of 
this process is further reduced by the overlap of these orbitals 
with the localized Cu one.
Moreover, such  virtual processes are insensitive to the flux. 
Any $\phi$ dependence arises from third-order corrections in which 
the $A_{1}$ particle virtually  goes around the trip clockwise or anticlockwise. 
In the ground state, of course, it chooses the wise in such a way to 
gain energy from the magnetic field. This is why a paramagnetic 
dependence on the flux is seen in Fig.(\ref{plot7}) and the correction goes 
like $-\phi^{2}$ at small $\phi$. 
This is interesting because it shows 
how the local symmetry can hinder the tunneling of bound pairs carrying 
conserved  quantum numbers; SFQ is not a necessary consequence 
of superconductivity if the pairs are not totalsymmetric.

In conclusion, we have found that Hubbard-like repulsive graphs  
hosting $W=0$ 2-body states can be  
used to model bound  pairs and their propagation.
Thus, SFQ may be found in purely repulsive 1$d$ Hubbard 
models only if the nodes are represented by a non-trivial basis.
For rings of $\cu$-units and weak O-O links we find a half-integer AB 
effect which is unambiguosly interpreted as SFQ. Next, we reported a 
counterexample, namely, the case of Cu-Cu links, when pairing is not 
leading to SFQ due to the large effective mass of the particles.
 By manipulating matrices whose dimension is the 
square root of the overall size of the Hilbert space we were able 
to exactly diagonalize the Hamiltonian: $\sqrt{m}\times\sqrt{m}$ 
matrices solve the $m\times m$ problem by means of a suitable 
disentanglement of the up and down spin configurations.

}

\end{document}